\documentstyle[epsf,aps,pra,multicol]{revtex}

\begin{document}
\draft
\title{Engineering two-mode interactions in ion traps}
\author{J.~Steinbach, J.~Twamley, and P.L.~Knight}
\address{{\it Optics Section, Blackett Laboratory, Imperial College,}\\
London SW7 2BZ, United Kingdom}
\date{April 8, 1997}
\maketitle

\begin{abstract}
We describe how two vibrational degrees of freedom of a single trapped ion
can be coupled through the action of suitably-chosen laser excitation. We
concentrate on a two-dimensional ion trap with dissimilar vibrational
frequencies in the $x$- and $y$-directions of motion, and derive from first
principles a variety of quantized two-mode couplings, concentrating on a
linear coupling which takes excitations from one mode to another. We
demonstrate how this can result in a state rotation, in which it is possible
to transfer the motional state of the ion from say the $x$-direction to the $%
y$-direction without prior knowledge of that motional state.
\end{abstract}


\begin{multicols}{2}

\section{Introduction}

In recent years advances in the cooling and trapping of ions have led to a
situation in which the centre-of-mass (CM) motion of trapped ions has to be
treated quantum mechanically \cite{lsr-cool}. This motion can be coherently
controlled by coupling the ion's external and internal degrees of freedom
through laser irradiation \cite
{blockley,cirac93,vogel95,even-odd,nonlinear-coherent}. Systems of trapped
ions have been employed to demonstrate experimentally the generation and
measurement of non-classical states of the ion's CM motion \cite
{meekhof96,leibfried96,monroe96,itano97}. Furthermore trapped ions have been
used to implement quantum logic gates \cite{monroe95,cirac95,james97}.

Most of the previous investigations have focussed on the one-dimensional
quantum motion of trapped ions. Recently, Gou {\it et.\ al.\/} \cite
{beam-splitter,pcs,pcas,intelligent,paircat} considered the generation of
particular two-mode states of an ion. In this paper we address the issue of
how to engineer a class of interactions between two of the quantized
motional degrees of freedom of a single trapped ion. We assume that the ion
is confined within a trap potential that can be closely approximated by a
two-dimensional harmonic well. In this case the CM motion of the ion is
completely equivalent to that of a two-dimensional harmonic oscillator,
characterized by two frequencies of oscillation $\nu_{\scriptstyle a}$ and $%
\nu_{\scriptstyle b}$ in orthogonal directions $x$ and $y,$ and the
corresponding operators $\left.\!\hat{a}^\dagger\!\right. (\hat{a})$ and $%
\left.\!\hat{b}^\dagger\!\right. (\hat{b})$ create (annihilate) vibrational
excitations in the $x$- and $y$-directions. The interaction that we want to
engineer is of the parametric form 
\begin{equation}
\hat{H}_{{\rm I}} = \hbar \left\{ g\,\left.\!\hat{a}^\dagger\!
\right.^{k_{\scriptstyle a}}
\hat{b}^{k_{\scriptstyle b}} + g^*\, \hat{a}^{k_{\scriptstyle a}} 
\left.\!\hat{b}^\dagger\!\right.^{k_{\scriptstyle b}}
\right\}\,,  \label{5}
\end{equation}
where ${k_{\scriptstyle a}}$ and ${k_{\scriptstyle b}}$ are positive
integers and $g$ is a complex coupling constant. In particular, we note that
the powers ${k_{\scriptstyle a}},$ and ${k_{\scriptstyle b}},$ can be
independently controlled to take on any positive integer numbers, and the
phase of the coupling constant $g,$ is freely adjustable. To give specific
examples of this class of interactions between the two vibrational modes $a$
and $b,$ we address the two coupling Hamiltonians, 
\begin{eqnarray}
\hat{H}_{{\rm I}}^{{\rm (1)}} &=& i\,\hbar g\,\left\{ \left.\!\hat{a}%
^\dagger\!\right. \hat{b} - \hat{a} \left.\!\hat{b}^\dagger\!\right.
\right\} \,,  \label{6} \\
\hat{H}_{{\rm I}}^{{\rm (3)}} &=& \hbar \left\{ g\,\left.\!\hat{a}%
^\dagger\!\right.^3 \hat{b} + g^*\, \hat{a}^3 \left.\!\hat{b}%
^\dagger\!\right. \right\} \,.  \label{7}
\end{eqnarray}
The Hamiltonian ({\ref{6}}) generates an active rotation of the
two-dimensional quantized motional state of the ion at a frequency $g,$
where $g$ is real. Here $\hat{H}_{{\rm I}}^{{\rm (1)}}$ is the kind of
Hamiltonian associated with a linear coupler or beam splitter in optics (see
e.g.\ \cite{lai91} and refs.\ therein). There, a photon in mode $a$ is
annihilated and a photon in mode $b$ is created, and vice versa. In a
trapped ion, vibrational anticorrelated SU(2) states of motion
characteristic of this kind of linear coupling can be generated \cite
{beam-splitter}. The linear coupling, $\hat{H}_{{\rm I}}^{{\rm (1)}},$ makes
it possible to transfer the motional state of the ion from, say the $x$%
-direction into the $y$-direction {\it without prior knowledge} of that
motional state and irrespective of whether it is a pure or a mixed state. In
the situation in which one may want to use the quantized motion in the $x$%
-direction for quantum computation \cite{monroe95,cirac95,james97}, perhaps
later entangling the quantum state of motion with internal electronic
states, the $y$-direction can then be employed as a quantum memory element.
Note the key point here is that states of motion in the $x$-direction can be
transferred entirely to the $y$-direction without reading out their nature,
entirely non-destructively. The Hamiltonian $\hat{H}_{{\rm I}}^{{\rm (3)}}$
is of the three-photon down-conversion kind: in optics, it represents a
process in which one pump photon in mode $b$ is annihilated and three
photons in mode $a$ are created, and vice versa. This process is known to be
highly peculiar: unlike its two-photon down-conversion cousin, quantization
of the pump is essential to avoid pathological divergences \cite{fisher84}.
These are avoided in a fully quantized treatment, where the pump and
down-converted field modes become highly entangled \cite{drobny92}.

In section \ref{generalraman}, we first introduce a two-mode Raman
transition which couples the electronic and motional degrees of freedom of
the ion. Choosing the initial state of the ion to be a direct product of an
arbitrary motional state and a specific electronic state, we then decouple
the electronic and motional dynamics of the ion through a particular
configuration of laser beams (section \ref{specificcoupling}). In the
Lamb-Dicke approximation and in the limit of suitable trap anisotropy we
obtain the above Hamiltonian ({\ref{5}}) for various sideband detunings of
the lasers. We then examine the severity of the approximations made to
obtain the Hamiltonian (\ref{5}). In section \ref{limits} we obtain
analytical estimates regarding the effects of off-resonant and higher
on-resonant processes. In section \ref{rotation} we specialize to the case ${%
k_{\scriptstyle a}} = {k_{\scriptstyle b}} =1,$ and show that the
Hamiltonian $\hat{H}_{{\rm I}}^{{\rm (1)}}$ rotates the motional quantum
state of the ion. Finally, we perform a numerical analysis of the complete
quantum dynamics and find that the Hamiltonian (\ref{6}) can be accurately
engineered over a range of parameters.

\section{General two-mode Raman coupling}

\label{generalraman}

In the following we describe the Raman coupling which we use to engineer the
Hamiltonian given in ({\ref{5}}). We consider an effective three-level ion
in a $\Lambda$-configuration, confined within a two-dimensional harmonic
trap as illustrated in Fig.\ref{fig1}. The trap is characterized by the two
frequencies $\nu_{\scriptstyle a}$ and $\nu_{\scriptstyle b}$ which describe
the harmonic potential in $x$- and $y$-direction respectively. As shown in
Fig.\ref{fig1}, the ion is excited by two linearly polarized laser beams
which propagate in the $x$- and $y$-directions connecting levels $|1\rangle
\Leftrightarrow |2\rangle$ and $|2\rangle \Leftrightarrow |3\rangle.$ These
beams are far detuned from the excited state $|2\rangle,$ in order to
generate a stimulated Raman transition between the two states $|1\rangle$
and $|3\rangle.$ We assume states $|1\rangle$ and $|3\rangle$ to be ground
state hyperfine sublevels.
\begin{minipage}[t]{0.48\textwidth}
\begin{figure}[t]
   {
   %
    \leavevmode
    \epsfxsize=85mm
    \epsffile{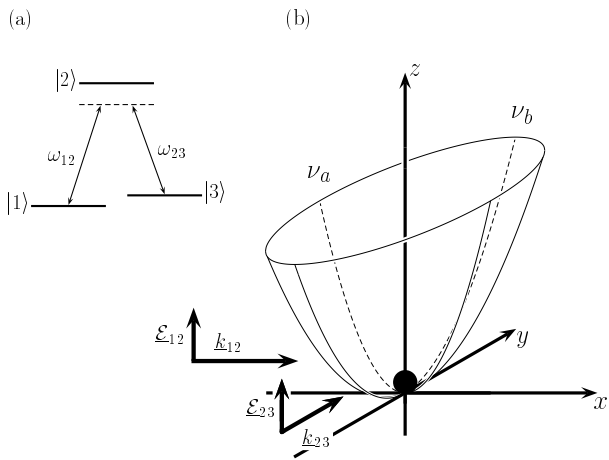}
   %
   }
\caption{
The two-mode Raman transition which couples the electronic and the
two motional degrees of freedom in the $x$- and $y$-direction. The effective
three-level ion shown in (a) is confined within a two-dimensional harmonic
trap. As illustrated in (b) two laser beams propagating in the $x$- and
$y$-directions generate a stimulated Raman transition between the ground states
$|1\rangle$ and $|3\rangle.$
}
\label{fig1}
\end{figure}
\end{minipage}

We do not include decoherence effects in our model for the following reason.
The Raman coupled energy level scheme greatly suppresses the spontaneous
emission between the two ground state levels $|3\rangle $ and $|1\rangle $
as these states are coupled by M1 and E2 transitions at best. At the same
time we neglect the effects of spontaneous emission from level $|2\rangle ,$
as the coupling to the excited state can be effectively eliminated over the
timescales of interest to us here when the laser beams are far detuned.
Another source of decoherence in ion trap experiments is classical noise in
the laser beams and trapping potential. This may be described using
so-called intrinsic decoherence models, e.g.\ \cite{moya-cessa93}, of
dephasing. The effects of this kind of decoherence have been seen in a
recent experiment by Meekhof et.\ al.\ \cite{meekhof96}. However, they
expect to significantly reduce decoherence from classical noise sources in
future experiments \cite{wineland97}. We thus do not include any decoherence
effects in our model.

Treating the laser excitations classically, the two electric fields are
described by 
\begin{eqnarray}
\underline{{\cal E}}_{12} (\hat{x},t) &=& \underline{e}_{12} \left\{
E_{12}\,e^{-i[k_{12}\hat{x} - \omega_{12} t]} + {\rm h.c.} \right\}\,, 
\nonumber \\
\underline{{\cal E}}_{23} (\hat{y},t) &=& \underline{e}_{23} \left\{
E_{23}\,e^{-i[k_{23}\hat{y} - \omega_{23} t]} + {\rm h.c.} \right\}\,,
\label{8}
\end{eqnarray}
where $\underline{e}_{12}$ and $\underline{e}_{23}$ are polarization
vectors, $k_{12}$ and $k_{23}$ are wavenumbers, and $\omega_{12}$ and $%
\omega_{23}$ are the frequencies of the lasers. We assume the laser phases
to be absorbed in the complex amplitudes $E_{12}$ and $E_{23}.$ In dipole
approximation this leads to the Hamiltonian 
\begin{eqnarray}
\hat{H} &=& \ \ \hbar \omega_1 |1\rangle \langle1| + \hbar \omega_2
|2\rangle \langle2| + \hbar \omega_3 |3\rangle \langle3|  \nonumber \\[2ex]
&& + \ \hbar \nu_{\scriptstyle a} (\left.\!\hat{a}^\dagger\!\right. \hat{a})
+ \hbar \nu_{\scriptstyle b} (\left.\!\hat{b}^\dagger\!\right. \hat{b}) 
\nonumber \\[2ex]
&& - \ \underline{D}_{12}. \underline{{\cal E}}_{12} - \underline{D}_{23}. 
\underline{{\cal E}}_{23} \,,  \label{10}
\end{eqnarray}
where we have denoted the dipole moments of the $|1\rangle \Leftrightarrow
|2\rangle$ and $|2\rangle \Leftrightarrow |3\rangle$ transitions by $%
\underline{D}_{12}$ and $\underline{D}_{23}$ respectively. The frequencies $%
\omega_1,\omega_2$ and $\omega_3,$ are associated with the energies of the
electronic states $|1\rangle,|2\rangle$ and $|3\rangle,$ and the operators $%
\hat{a} (\left.\!\hat{a}^\dagger\!\right.)$ and $\hat{b} (\left.\!\hat{b}%
^\dagger\!\right.)$ are the annihilation (creation) operators for
vibrational quanta in the $x$- and $y$-directions. These operators are
related to the position of the ion in the $x$-$y$ plane through 
\begin{eqnarray}
\hat{x} &=& \Delta x_0 \, (\hat{a} + \left.\!\hat{a}^\dagger\!\right.)\,, 
\nonumber \\
\hat{y} &=& \Delta y_0 \, (\hat{b} + \left.\!\hat{b}^\dagger\!\right.)\,,
\label{20}
\end{eqnarray}
where $\Delta x_0 = (\hbar / 2 \nu_{\scriptstyle a} m )^{1/2}$ and $\Delta
y_0 = (\hbar / 2 \nu_{\scriptstyle b} m )^{1/2}$ are the widths of the
ground state in the two-dimensional harmonic oscillator potential in $x$-
and $y$-directions, and $m$ is the mass of the ion. If the laser beams are
sufficiently far detuned, i.e. 
\begin{equation}
|\Delta_{12}|, |\Delta_{23}| \gg |g_{12}|, |g_{23}|, |\Delta_{12} -
\Delta_{23}|\,,  \label{30}
\end{equation}
the two ground states $|1\rangle$ and $|3\rangle$ are coupled via a
stimulated Raman transition and the excited state $|2\rangle$ can be
adiabatically eliminated. In the above inequality we have defined the laser
detunings $\Delta_{12} = (\omega_2 -\omega_1) - \omega_{12}, \ \Delta_{23} =
(\omega_2 - \omega_3) - \omega_{23},$ and the dipole coupling constants $%
g_{12} = \langle1| \underline{D}_{12}.\underline{e}_{12}
|2\rangle\,E_{12}/\hbar, \ g_{23} = \langle3| \underline{D}_{23}.\underline{e%
}_{23} |2\rangle\,E_{23}/\hbar.$ As described in Appendix A, the adiabatic
elimination procedure leads to the Hamiltonian 
\begin{eqnarray}
\hat{H} &=& \ \ \hbar \widetilde{\omega}_1 \, |1\rangle \langle1| + \hbar 
\widetilde{\omega}_3 \, |3\rangle \langle3|  \nonumber \\[1ex]
&& + \hbar \nu_{\scriptstyle a} \, (\left.\!\hat{a}^\dagger\!\right. \hat{a}%
) + \hbar \nu_{\scriptstyle b} \, (\left.\!\hat{b}^\dagger\!\right. \hat{b})
\nonumber \\[1ex]
&& - \hbar g_{13} \, e^{-i\left[k_{12}\hat{x} - k_{23}\hat{y} - (\omega_{12}
- \omega_{23})t \right]} \otimes |1\rangle \langle3|  \nonumber \\[1ex]
&& - \hbar g_{13}^* \, e^{i \left[k_{12}\hat{x} - k_{23}\hat{y} -
(\omega_{12} - \omega_{23})t \right]} \otimes |3\rangle \langle1| \,,
\label{40}
\end{eqnarray}
where we have dropped the term describing the free energy of the excited
state $|2\rangle$ as in the far detuned limit ({\ref{30}}) the excited state
is no longer connected to the two ground states. Furthermore, we have
defined the Raman coupling constant 
\begin{equation}
g_{13} = g_{12} g_{23}^*\, \left( \frac{1}{\Delta_{12}} + \frac{1}{%
\Delta_{23}} \right)\,,  \label{50}
\end{equation}
and the energies $\hbar \widetilde{\omega}_1$ and $\hbar \widetilde{\omega}%
_3 $ of the ground state levels $|1\rangle$ and $|3\rangle,$ which are Stark
shifted as a result of the adiabatic elimination of the excited state, are 
\begin{eqnarray}
\widetilde{\omega}_1 &=& \omega_1 - \frac{2 | g_{12} |^2}{\Delta_{12}}\,, 
\nonumber \\
\widetilde{\omega}_3 &=& \omega_3 - \frac{2 | g_{23} |^2}{\Delta_{23}}\,.
\label{55}
\end{eqnarray}
In order to proceed, we will consider the Raman coupling Hamiltonian ({\ref
{40}}) in the interaction picture of $\hat{H}_{{\rm 0}} = \hbar \widetilde{%
\omega}_1 \, |1\rangle \langle1| + \hbar \widetilde{\omega}_3 \, |3\rangle
\langle3| + \hbar \nu_{\scriptstyle a} \, (\left.\!\hat{a}^\dagger\!\right. 
\hat{a}) + \hbar \nu_{\scriptstyle b} \, (\left.\!\hat{b}^\dagger\!\right. 
\hat{b}),$ and transform to the new Hamiltonian 
\begin{equation}
\hat{H}_{{\rm I}} = \,e^{i\,\hat{H}_{{\rm 0}} t\,/\hbar} \left( \hat{H} - 
\hat{H}_{{\rm 0}} \right) \,e^{-i\,\hat{H}_{{\rm 0}} t\,/\hbar}\,.
\label{58}
\end{equation}
In doing so and replacing the position operators $\hat{x}$ and $\hat{y}$ by (%
{\ref{20}}) we obtain the interaction Hamiltonian 
\begin{eqnarray}
\hat{H}_{{\rm I}} &=& - \hbar g_{13} \ \exp\left[- \frac{1}{2} ( \eta_{12}^2
+ \eta_{23}^2 ) \right] \ |1\rangle \langle3|  \nonumber \\[1.5ex]
&& \otimes \sum_{m,\mu,n,\nu} \frac{(-i \eta_{12})^{m+\mu}}{m!\, \mu!} \frac{
(i\eta_{23})^{n+\nu}} {n!\, \nu!}\, \left.\!\hat{a}^\dagger\!\right.^m
a^\mu\, \left.\!\hat{b}^\dagger\!\right.^\nu b^n  \nonumber \\[1.5ex]
&& \times \, \exp \Big [\,i\,\big ( \nu_{\scriptstyle a} [m-\mu] + \nu_{%
\scriptstyle b} [\nu-n] + \Delta_{13} \big )\,t\,\Big ] \ + \ {\rm h.c.}\,,
\nonumber \\[1.5ex]
&& \label{60}
\end{eqnarray}
where we have defined the Raman detuning, 
\begin{equation}
\Delta_{13} = \omega_{12} - \omega_{23} - (\widetilde{\omega}_3 - \widetilde{%
\omega}_1)\,,  \label{70}
\end{equation}
and the Lamb-Dicke parameters in $x$- and $y$-direction, $\eta_{12} = \Delta
x_0\,k_{12},$ and $\eta_{23} = \Delta y_0\,k_{23}.$ The square of the
Lamb-Dicke parameter gives the ratio of the single photon recoil energy to
the energy level spacing in the harmonic oscillator potential.

\section{Specific coupling scheme}

\label{specificcoupling}

In this section we construct a particular configuration of Raman lasers to
decouple the electronic and motional dynamics of the trapped ion for
suitably-chosen initial electronic states. This is done by symmetrically
combining two Raman transitions as described below. We then obtain the
Hamiltonian ({\ref{5}}) in the Lamb-Dicke approximation and in the limit of
suitable trap anisotropy for specific sideband detunings of the lasers.

The electronic and motional dynamics can be decoupled in general for the
Hamiltonian $\hat{H}_{{\rm I}} = \hat{M} \otimes |1\rangle \langle3| + \hat{M%
}^\dagger \otimes |3\rangle \langle1|,$ where $\hat{M}$ may be any operator
that acts on the motional degrees of freedom only. This is done through the
addition of another interaction generated by $\hat{H}_{{\rm I}}^\prime = 
\hat{M} \otimes |3\rangle \langle1| + \hat{M}^\dagger \otimes |1\rangle
\langle3|.$ Combining both interactions, we have $\hat{H}_{{\rm I}}^{{\rm tot%
}} = \hat{H}_{{\rm I}} + \hat{H}_{{\rm I}}^\prime,$ so that the combined
Hamiltonian, $\hat{H}_{{\rm I}}^{{\rm tot}} = (\hat{M} + \hat{M}^\dagger)
\otimes (|3\rangle \langle1| + |1\rangle \langle3|),$ factorizes. For the
case where $\hat{H}_{{\rm I}}$ is given by ({\ref{60}}), $\hat{H}_{{\rm I}%
}^\prime$ can be generated by an extra pair of Raman lasers with suitable
detunings, propagation directions and phases. To be more specific, we
require a symmetric combination of two Raman transitions, so that 
\begin{eqnarray}
\Delta_{13}^\prime &=& - \Delta_{13}\,,  \label{76} \\[1.5ex]
\eta_{12}^\prime &=& - \eta_{12}\,,  \nonumber \\
\eta_{23}^\prime &=& - \eta_{23}\,,  \label{77} \\[1.5ex]
g_{13}^\prime &=& g_{13}^*\,,  \label{78}
\end{eqnarray}
where all quantities without primes correspond to the first pair of Raman
lasers and all primed quantities refer to the second pair. If, for the first
pair of lasers, the Raman detuning $\Delta_{13}$ is given by ({\ref{70}}),
then the first condition ({\ref{76}}), requires an appropriate choice of the
frequencies $\omega_{12}^\prime$ and $\omega_{23}^\prime$ for the second
pair, so that $\Delta_{13}^\prime = \omega_{12}^\prime - \omega_{23}^\prime
- (\widetilde{\omega}_3 - \widetilde{\omega}_1) = - \Delta_{13}.$ This is
illustrated in Fig.\ref{fig2}. The second condition ({\ref{77}}), is
satisfied by choosing the second pair of beams to be counter-propagating
with respect to the first pair, so that $k_{12}^\prime = - k_{12},$ and $%
k_{23}^\prime = - k_{23},$ as seen from the definition of the Lamb-Dicke
parameters, $\eta_{12} = \Delta x_0\,k_{12},$ and $\eta_{23} = \Delta
y_0\,k_{23}.$ Here, we have neglected the differences $|k_{12}| -
|k_{12}^\prime|,$ and $|k_{23}| - |k_{23}^\prime|,$ since $|\omega_{12} -
\omega_{12}^\prime| \ll \omega_{12} , \omega_{12}^\prime,$ and $|\omega_{23}
- \omega_{23}^\prime| \ll \omega_{23} , \omega_{23}^\prime.$ This
restriction can be lifted if one chooses the second pair of lasers to be not
exactly counter-propagating with the first. The third condition ({\ref{78}}%
), requires a suitable choice of laser phases for the two pairs of Raman
beams which can be easily read from ({\ref{50}}). The symmetric combination
of the two Raman transitions as specified by ({\ref{76}})--({\ref{78}}) then
leads to the interaction Hamiltonian %
\end{multicols}
\begin{eqnarray}
\hat{H}_{{\rm I}}^{{\rm tot}} &=& - \hbar \left\{ g_{13} \ \exp\left[- \frac{%
1}{2} ( \eta_{12}^2 + \eta_{23}^2 ) \right] \right. \sum_{m,\mu,n,\nu} \frac{%
(-i \eta_{12})^{m+\mu}}{m!\, \mu!} \frac{(i\eta_{23})^{n+\nu}} {n!\, \nu!}\,
\left.\!\hat{a}^\dagger\!\right.^m a^\mu\, \left.\!\hat{b}%
^\dagger\!\right.^\nu b^n  \nonumber \\[2ex]
&& \times \, \exp \Big[\,i\, ( \nu_{\scriptstyle a} [m-\mu] + \nu_{%
\scriptstyle b} [\nu-n] + \Delta_{13}) \,t\,\Big] \ + \ {\rm h.c.} \bigg \} %
\otimes \Big \{ |1\rangle\langle3| + |3\rangle\langle1| \Big \} \,,
\label{80}
\end{eqnarray}
which factorizes. We now assume the ion to be initially in a direct product
of its motional and electronic state with the electronic state prepared as $%
|+\rangle = (|1\rangle + |3\rangle)/\sqrt{2}.$ This superposition state $%
|+\rangle,$ can be prepared from the ground state $|1\rangle,$ by applying a
resonant $\pi/2$-pulse $(\Delta_{13}=0)$, if the ion is confined within the
Lamb-Dicke limit, i.e.\ $\eta_{12}, \eta_{23} \ll 1,$ \cite{monroe96}. The
dynamics generated by eq.({\ref{80}}) acting on this state factors and
leaves the electronic state unchanged. This allows us to reduce the dynamics
to that of the motional degrees of freedom only, and we write 
\begin{eqnarray}
\hat{H}_{{\rm I}}^{{\rm tot}} &=& - \hbar g_{13} \ \exp\left[- \frac{1}{2} (
\eta_{12}^2 + \eta_{23}^2 ) \right] \quad \sum_{m,\mu,n,\nu} \frac{(-i
\eta_{12})^{m+\mu}}{m!\, \mu!} \frac{(i\eta_{23})^{n+\nu}} {n!\, \nu!}\,
\left.\!\hat{a}^\dagger\!\right.^m a^\mu\, \left.\!\hat{b}%
^\dagger\!\right.^\nu b^n  \nonumber \\[2ex]
&& \times \, \exp \Big [\,i\,( \nu_{\scriptstyle a} [m-\mu] + \nu_{%
\scriptstyle b} [\nu-n] + \Delta_{13})\,t\,\Big] \ + \ {\rm h.c.}  \label{90}
\end{eqnarray}
\hspace*{\fill}\rule[0.4pt]{0.4pt}{\baselineskip}%
\rule[\baselineskip]{0.5\textwidth}{0.4pt}
\begin{multicols}{2}
\noindent
\begin{minipage}[t]{0.48\textwidth}
\begin{figure}[t]
{
  \leavevmode
  \epsfxsize=100mm
  \epsffile{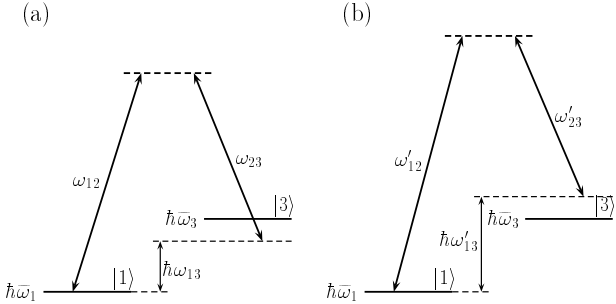}
 }
\caption{ Schematic diagram of two symmetric Raman transitions which in
combination decouple the electronic and motional dynamics of the trapped ion
for suitably-chosen initial electronic states. In (a), the frequencies $\omega_{12}$
and $\omega_{23}$ of the two lasers that generate the stimulated Raman
transition between the ground state levels $|1\rangle$ and $|3\rangle$ are
chosen such that the Raman detuning $\Delta_{13} = \omega_{12} - \omega_{23}
- (\widetilde{\omega}_3 - \widetilde{\omega}_1)$ is positive. The coupling
lasers are red detuned with respect to the $|1\rangle \Leftrightarrow
|3\rangle$ transition. In (b), we show the symmetric Raman transition to
(a). The frequencies $\omega_{12}^\prime$ and $\omega_{23}^\prime$ of the
coupling lasers are adjusted so that $\Delta_{13}^\prime =
\omega_{12}^\prime - \omega_{23}^\prime - (\widetilde{\omega}_3 - \widetilde{%
\omega}_1) = - \Delta_{13}.$ The coupling beams are blue detuned with
respect to the $|1\rangle \Leftrightarrow |3\rangle$ transition. For the two
transitions to be symmetric we additionally require the coupling beams in
(b) to be counter-propagating with respect to the beams in (a). }
\label{fig2}
\end{figure}
\end{minipage}
\vspace*{\baselineskip}

We now discuss the sideband detunings which, in the Lamb-Dicke approximation
and in the limit of suitable trap anisotropy, lead to the desired
interaction ({\ref{5}}). In particular, detuning the two pairs of Raman
lasers to specific vibrational sidebands allows us to choose specific values
for ${k_{\scriptstyle a}}$ and ${k_{\scriptstyle b}}$ in ({\ref{5}}). Since
we require the two Raman transitions to be symmetric, it is sufficient to
consider the first pair of Raman lasers. Therefore, we return to the
vibronic Raman coupling Hamiltonian ({\ref{60}}). From eq.({\ref{60}}) it is
clear that by fixing the size of the detuning $\Delta_{13},$ i.e.\ by
choosing the frequencies of the two coupling lasers, we can tune to a
resonance between specific vibronic levels. As illustrated in Fig.\ref{fig3}
we introduce a virtual level $|c\rangle$ with energy $\hbar \omega_c$ to
help visualize the Raman transitions between the ion's vibronic levels. If
we let $\omega_{12} = (\omega_c - \widetilde{\omega}_1) - k_{\scriptstyle a}
\nu_{\scriptstyle a},$ and $\omega_{23} = (\omega_c - \widetilde{\omega}_3)
- k_{\scriptstyle b} \nu_{\scriptstyle b},$ then with respect to level $%
|c\rangle,$ the first laser is tuned to the ${k_{\scriptstyle a}}$-th red
sideband of the ion's vibration in the $x$-direction, the second laser is
tuned to the ${k_{\scriptstyle b}}$-th red sideband of the vibration in the $%
y$-direction and the Raman detuning is 
\begin{equation}
\Delta_{13} = {k_{\scriptstyle b}} \nu_{\scriptstyle b} - {k_{\scriptstyle a}%
} \nu_{\scriptstyle a}\,.  \label{100}
\end{equation}
This situation is illustrated in Fig.\ref{fig3} for the specific example, ${%
k_{\scriptstyle a}} = {k_{\scriptstyle b}} = 1.$ Now, if only on-resonant
terms in ({\ref{60}}) are retained, we have $m = \mu + {k_{\scriptstyle a}},$
and $n = \nu + {k_{\scriptstyle b}},$ and we obtain the Hamiltonian 
\begin{equation}
\hat{H}_{{\rm I}} = |1\rangle \langle3| \otimes \sum_{\mu,\nu} \hbar
g(\mu,\nu)\, \left.\!\hat{a}^\dagger\!\right.^{k_{\scriptstyle a}}
\left.\!\hat{a}^\dagger\!\right.^\mu\hspace{-4pt}\hat{a}^\mu\,
\left.\!\hat{b}^\dagger\!\right.^\nu\hspace{-3pt}\hat{b}^%
\nu\hspace{2pt}\hat{b}^{k_{\scriptstyle b}} \ + \ 
{\rm h.c.}\,,  \label{110}
\end{equation}
where we have defined the coupling constants 
\begin{eqnarray}
g(\mu,\nu) &=& - g_{13} \ \exp\left[- \frac{1}{2} ( \eta_{12}^2 +
\eta_{23}^2 ) \right]  \nonumber \\[2ex]
&& \times \frac{(-i \eta_{12})^{2 \mu + {k_{\scriptstyle a}}}}{\mu!\, (\mu + 
{k_{\scriptstyle a}})!} \frac{(i\eta_{23})^{2 \nu + {k_{\scriptstyle b}}}}{%
\nu!\, (\nu + {k_{\scriptstyle b}})!}\,.  \label{115}
\end{eqnarray}
This is a two-mode generalization of the non-linear Jaynes-Cummings model
introduced by Vogel {\it et.\ al.\/} \cite{vogel95}. It is important for the
trap frequencies $\nu_{\scriptstyle a}$ and $\nu_{\scriptstyle b}$ to be
non-commensurate to arrive at this result. This becomes clear from Fig.\ref
{fig3}. 
\begin{minipage}[t]{0.48\textwidth}
\begin{figure}[t]
   {
    \leavevmode
    \epsfxsize=95mm
    \epsffile{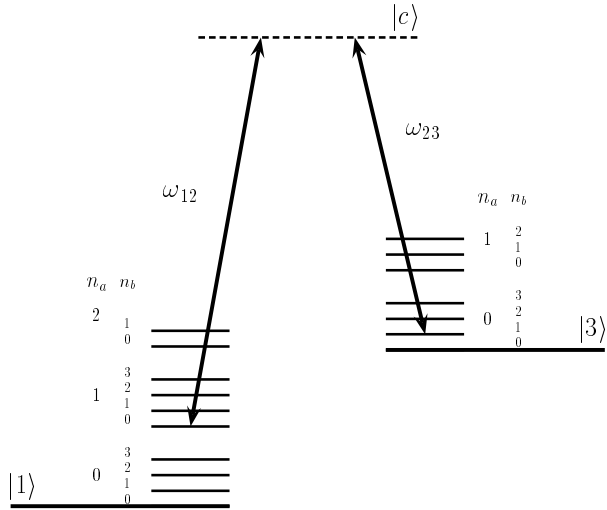}
   }
\caption{ Schematic diagram of the vibronic energy levels that are connected
by the two laser beams which generate the stimulated Raman transition.
The two Raman lasers are tuned such that the Raman detuning $\Delta_{13} = %
{k_{\scriptstyle b}} \nu_{\scriptstyle b} - {k_{\scriptstyle a}} %
\nu_{\scriptstyle a},$ with $k_{\scriptstyle a} = {k_{\scriptstyle b}} = 1.$
With respect to the virtual level $|c\rangle,$ the laser propagating in the
$x$-direction is tuned to the first red sideband of the ion's vibration in the
$x$-direction and the laser that propagates in the $y$-direction is tuned to the
first red sideband of the ion's vibration in the $y$-direction. This causes a
resonant transition between the vibronic states $|{n_a - 1}\rangle_{
\scriptstyle a}|{n_b}\rangle_{\scriptstyle b}|{3}\rangle \Leftrightarrow |{%
n_a}\rangle_{\scriptstyle a}|{n_b - 1}\rangle_{\scriptstyle b}|{1}\rangle,$
where the states $|{n_a}\rangle_{\scriptstyle a}|{n_b}\rangle_{\scriptstyle %
b}$ denote the usual number state basis for the two-dimensional harmonic
oscillator and the numbers $n_a$ and $n_b$ give the number of vibrational
excitations in the $x$- and $y$-direction respectively. }
\label{fig3}
\end{figure}
\end{minipage}
\vspace{\baselineskip}

\noindent
If the trapping potential is isotropic, $\nu_{\scriptstyle a} = \nu_{%
\scriptstyle b},$ and the energy levels become degenerate. Consequently, the
Raman transition Hamiltonian ({\ref{60}}) contains on-resonant terms in
addition to the ones retained in ({\ref{110}}). In the example ${k_{%
\scriptstyle a}} = {k_{\scriptstyle b}} = 1,$ this leads to a coupling
Hamiltonian $\hat{H}_{{\rm I}} \propto [1 + \eta^2\,(\left.\!\hat{a}%
^\dagger\!\right. \hat{b} + \hat{a} \left.\!\hat{b}^\dagger\!\right. -
\left.\!\hat{a}^\dagger\!\right. \hat{a} - \left.\!\hat{b}^\dagger\!\right. 
\hat{b}) + O(\eta^4)]\otimes |1\rangle \langle3| + {\rm h.c.}\,,$ where we
have assumed the Lamb-Dicke parameters to be of the same order of magnitude, 
$\eta_{12} \approx \eta_{23} \approx \eta.$ In general, if the frequencies $%
\nu_{\scriptstyle a}$ and $\nu_{\scriptstyle b}$ are commensurate the Raman
transition Hamiltonian ({\ref{60}}) contains resonances in addition to the
ones considered in eq.({\ref{110}}). As we will show in the next section, in
the Lamb-Dicke limit, the coupling constants corresponding to these
additional resonances can be greatly reduced by increasing the ratio of the
trap frequencies, $\nu_{\scriptstyle a}/\nu_{\scriptstyle b}.$

The symmetric Raman transition ({\ref{110}}) is generated by a second pair
of lasers as specified in eqs.({\ref{76}})--({\ref{78}}). In particular, we
note that ({\ref{76}}) can be satisfied with the choice $\omega
_{12}^{\prime }=(\omega _{c}-\widetilde{\omega }_{1})+{k_{\scriptstyle a}}%
\nu _{\scriptstyle a},$ and $\omega _{23}^{\prime }=(\omega _{c}-\widetilde{%
\omega }_{3})+{k_{\scriptstyle b}}\nu _{\scriptstyle b},$ for the
frequencies of the second pair of lasers. With respect to the virtual level $%
|c\rangle ,$ these lasers are then detuned by the same amount as the first
pair but to the blue vibrational sidebands rather than the red. Combining
both Raman transitions, we obtain the reduced Hamiltonian 
\begin{equation}
\hat{H}_{{\rm I}}^{{\rm tot}}=\sum_{\mu ,\nu }\hbar g(\mu ,\nu )\,\left. \!%
\hat{a}^{\dagger }\!\right. ^{{k_{\scriptstyle a}}}\left. \!\hat{a}^{\dagger
}\!\right. ^{\mu }\hspace{-4pt}\hat{a}^{\mu }\,\left. \!\hat{b}^{\dagger
}\!\right. ^{\nu }\hspace{-3pt}\hat{b}^{\nu }\hspace{2pt}\hat{b}^{{k_{%
\scriptstyle b}}}\ +\ {\rm h.c.}\,,  \label{130}
\end{equation}
for the motional dynamics of the trapped ion as discussed above.

In the last step, we now assume the Lamb-Dicke limit, where $\eta_{12},
\eta_{23} \ll 1.$ In this limit we approximate eq.({\ref{130}}) by keeping
only the lowest order terms in $\eta_{12}$ and $\eta_{23}.$ From eq.({\ref
{115}}), these are the terms $\mu = \nu = 0,$ and we obtain 
\begin{equation}
\hat{H}_{{\rm I}}^{{\rm tot}} = \hbar \left\{ g\,\left.\!\hat{a}%
^\dagger\!\right.^{k_{\scriptstyle a}} \hat{b}^{k_{\scriptstyle b}} + g^*\, 
\hat{a}^{k_{\scriptstyle a}} \left.\!\hat{b}%
^\dagger\!\right.^{k_{\scriptstyle b}} \right\}\,,  \label{140}
\end{equation}
where $g = g(0,0),$ is given in ({\ref{115}}). The above Hamiltonian ({\ref
{140}}) realizes the desired interaction ({\ref{5}}), between the two modes $%
a$ and $b$ of the ion's motion in the $x$- and $y$-direction.
We note that the coupling constant $g$ depends on the Lamb-Dicke parameters
through the factor $\eta_{12}^{k_{\scriptstyle a}} 
\eta_{23}^{k_{\scriptstyle b}}.$ Consequently for fixed laser power,
i.e.\ fixed $|g_{12}|$ and $|g_{23}|,$ and small Lamb-Dicke parameters,
the coupling strength may be very small. One can increase the coupling 
constant $g$ by increasing the laser power while at the same time maintaining
inequality (\ref{30}). This permits us to ignore the spontaneous emission 
from the excited state $|2\rangle$ on a timescale
\begin{equation}
 T \ll T_{\rm spont} = \left( \frac{|g_{12}|^2}{\Delta_{12}^2} + 
 \frac{|g_{23}|^2}{\Delta_{23}^2} \right)^{-1} \gamma^{-1}\,,  
 \label{141}
\end{equation}
where $\gamma$ is the rate of spontaneous decay from level $|2\rangle$
\cite{plenio}. This is important as the decoupling of the motional and
electronic dynamics relies on maintaining the coherence of the electronic
degrees of freedom. In  section \ref{rotation} we will compare the
timescales for spontaneous emission and the Raman-generated motional dynamics
for the specific case of rotation (\ref{6}), given the parameters of 
recent experiments \cite{meekhof96}.

\section{Limitations}

\label{limits}

In this section we further discuss the approximations under which the
Hamiltonian ({\ref{140}}), gives a valid description of the system dynamics.
First, we address the size of the corrections that we have neglected in the
Lamb-Dicke approximation. We then show that the coupling constants of the
additional resonances in the case of commensurate trap frequencies can be
made as small as these corrections for a suitably large ratio of the trap
frequencies, $\nu_{\scriptstyle a}/\nu_{\scriptstyle b}.$ Finally, we
discuss the limitations imposed on our Hamiltonians from neglecting
off-resonant transitions.

\subsection{Lamb-Dicke approximation}

From the previous section it is clear that the Lamb-Dicke limit is an
important requirement for us to engineer the desired interaction ({\ref{140}}%
). The Lamb-Dicke approximation led us from ({\ref{130}}) to ({\ref{140}})
under the assumption $\eta_{12}, \eta_{23} \ll 1.$ We note that both ({\ref
{130}}) and ({\ref{140}}) couple the same vibrational states 
\begin{equation}
|{m}\rangle_{\scriptstyle a}|{n+{k_{\scriptstyle b}}}\rangle_{\scriptstyle %
b} \Longleftrightarrow |{m+{k_{\scriptstyle a}}}\rangle_{\scriptstyle a}|{n}%
\rangle_{\scriptstyle b}\,,  \label{170}
\end{equation}
where $|{m}\rangle_{\scriptstyle a}|{n}\rangle_{\scriptstyle b},$ denotes
the usual number state basis for the two-dimensional harmonic oscillator.
Therefore, we do not neglect any additional resonances between other states
than the ones given in ({\ref{170}}) by making the Lamb-Dicke approximation.

We define the Lamb-Dicke approximation for suitably small $\eta_{12}$, $%
\eta_{23}$ to be the approximation where all terms in ({\ref{130}}) of order 
$\eta^2$ smaller than the leading term are neglected i.e. 
\begin{equation}
\frac{|g(\mu,\nu)|}{|g(0,0)|} \leq O(\eta^2)\,,  \label{180}
\end{equation}
where we have assumed the Lamb-Dicke parameters to be of the same order of
magnitude, $\eta_{12} \approx \eta_{23} \approx \eta.$

It is important to note that the orthogonality of the Raman laser beams
shown in Fig.\ref{fig1} is not essential. In fact, the size of the
Lamb-Dicke parameters can be reduced by changing the geometry of the lasers
and choosing the two Raman beams to be almost counter-propagating. In this
situation the wave vectors $\underline{k}_{12}$ and $\underline{k}_{23}$ of
the two Raman beams have to be added and the numbers $k_{12}$ and $k_{23}$
in ({\ref{40}}) are then the projections of $\underline{k} = \underline{k}%
_{12} + \underline{k}_{23}$ onto the $x$- and $y$-axes respectively.

\subsection{Trap anisotropy}

\label{anisotropy}

As we have mentioned in the previous section, even in the case of an
anisotropic trap, there are on-resonant terms in addition to the ones
included in ({\ref{110}}) when the trap frequencies are commensurate. This
is illustrated in Fig.\ref{fig4}, where $\nu_{\scriptstyle a} = 5\,\nu_{%
\scriptstyle b},$ and again, ${k_{\scriptstyle a}} = {k_{\scriptstyle b}} =
1.$ In addition to the $|{m-1}\rangle_{\scriptstyle a}|{n}\rangle_{%
\scriptstyle b}|{3}\rangle \Leftrightarrow |{m}\rangle_{\scriptstyle a}|{n-1}%
\rangle_{\scriptstyle b}|{1}\rangle$ transition shown in Fig.\ref{fig3}, the 
$|{m}\rangle_{\scriptstyle a}|{n-4}\rangle_{\scriptstyle b}|{3}\rangle
\Leftrightarrow |{m}\rangle_{\scriptstyle a}|{n}\rangle_{\scriptstyle b}|{1}%
\rangle$ transition is resonantly coupled as in Fig.\ref{fig4}.  
In the
following we show that in the Lamb-Dicke limit, the coupling constants $%
\widetilde{g},$ corresponding to these additional resonances satisfy
\begin{equation}
\frac{|\widetilde{g}|}{|g(0,0)|} \leq O(\eta^2)\,,  \label{3990}
\end{equation}
if the ratio of the trap frequencies is chosen large enough. These
additional terms can thus be neglected in the Lamb-Dicke approximation.
\begin{minipage}[t]{0.48\textwidth}
\begin{figure}[t]
   {   
    \leavevmode
    \epsfxsize=95mm
    \epsffile{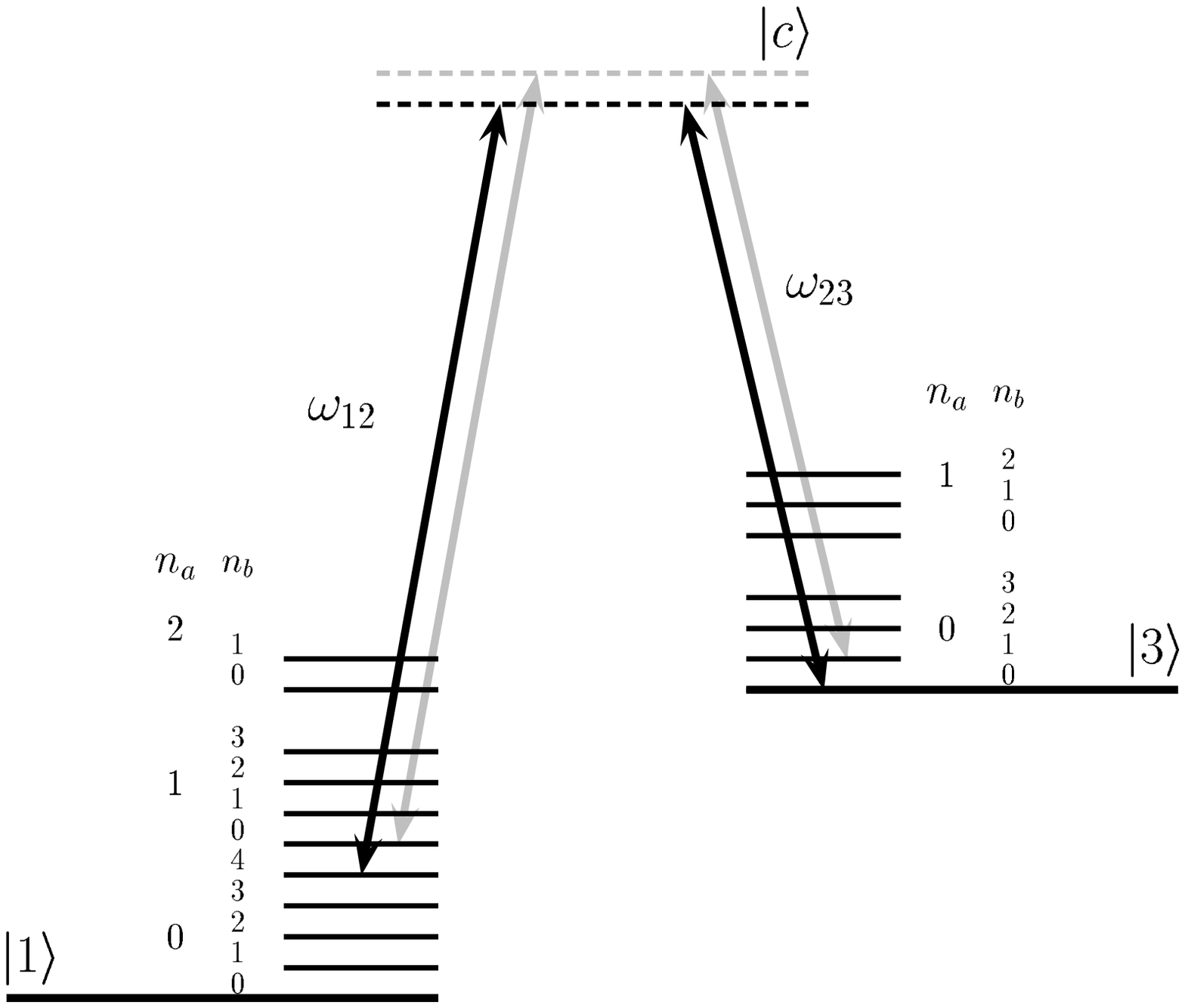}
   }   
\caption{
   Vibronic energy level diagram for the case of the two trap frequencies
   $\nu_{\scriptstyle a}$ and $\nu_{\scriptstyle b}$ being multiples of each
   other, $\nu_{\scriptstyle a} = 5\,\nu_{\scriptstyle b}.$ As in Fig.\ref{fig3}
   the two Raman lasers are tuned such that the Raman detuning $\Delta_{13} =
   k_{\scriptstyle b} \nu_{\scriptstyle b} - k_{\scriptstyle a}
   \nu_{\scriptstyle a},$ with $k_{\scriptstyle a} = k_{\scriptstyle b} = 1.$
   In addition to the  desired resonant transition
   $|{n_a - 1}\rangle_{\scriptstyle a}|{n_b}\rangle_{\scriptstyle b}
   |{3}\rangle \Leftrightarrow |{n_a}\rangle_{\scriptstyle a}|{n_b - 1}
   \rangle_{\scriptstyle b}|{1}\rangle,$ shown in grey, the $|{n_a}
   \rangle_{\scriptstyle a}|{n_b-4}\rangle_{\scriptstyle b}|{3}\rangle
   \Leftrightarrow |{n_a}\rangle_{\scriptstyle a}|{n_b}
   \rangle_{\scriptstyle b}|{1}\rangle$ transition is resonantly coupled as
   shown in black. In the Lamb-Dicke limit, the coupling constant corresponding
   to this additional resonance can be reduced to the size of the corrections
   to the Lamb-Dicke approximation for the desired resonance by
   increasing the ratio of the trap frequencies, $\nu_{\scriptstyle a}/
   \nu_{\scriptstyle b}.$
}
\label{fig4}
\end{figure}
\end{minipage}
\vspace{\baselineskip}

We start by deriving the resonances that occur if the two trap frequencies $%
\nu _{\scriptstyle a}$ and $\nu _{\scriptstyle b}$ are multiples of each
other. Without loss of generality we choose 
\begin{equation}
\nu _{\scriptstyle a}=l\,\nu _{\scriptstyle b}\,,  \label{4000}
\end{equation}
where $l$ is a positive integer number. In deriving the interaction ({\ref
{140}}) the laser frequencies were chosen to give the detunings $\Delta
_{13}={k_{\scriptstyle b}}\nu _{\scriptstyle b}-{k_{\scriptstyle a}}\nu _{%
\scriptstyle a},$ and $\Delta _{13}^{\prime }={k_{\scriptstyle a}}\nu _{%
\scriptstyle a}-{k_{\scriptstyle b}}\nu _{\scriptstyle b},$ for the two
pairs of coupling beams respectively. We will explicitly consider only the
first of these two cases, i.e. $\Delta _{13}={k_{\scriptstyle b}}\nu _{%
\scriptstyle b}-{k_{\scriptstyle a}}\nu _{\scriptstyle a},$ since the second
follows analogously by interchanging the operators $\hat{a}\Leftrightarrow
\left. \!\hat{a}^{\dagger }\!\right. ,$ and $\hat{b}\Leftrightarrow \left. \!%
\hat{b}^{\dagger }\!\right. ,$ and leads to the same limits for the trap
ratio $l=\nu _{\scriptstyle a}/\nu _{\scriptstyle b}.$ Now, with ({\ref{4000}%
}), the resonance condition in ({\ref{60}}) becomes 
\begin{equation}
-l\,(\mu -m)-(n-\nu )-l\,{k_{\scriptstyle a}}+{k_{\scriptstyle b}}=0\,,
\label{4010}
\end{equation}
where all numbers are positive integers. In order to simplify the discussion
we categorize the resonances by introducing an integer number $N,$ and
rewrite ({\ref{4010}}), so that 
\begin{eqnarray}
\mu -m &=&-{k_{\scriptstyle a}}+N\,,  \nonumber \\
\nu -n &=&-{k_{\scriptstyle b}}+l\,N\,.  \label{4020}
\end{eqnarray}
Following this categorization, we divide the resonances determined by ({\ref
{4010}}) into the three cases, [{\it 1\/}] $N=0,$ [{\it 2\/}] $N$ positive,
and [{\it 3\/}] $N$ negative. We subdivide case [{\it 2\/}] further into 
\begin{eqnarray}
\lbrack {\it 2a\/}]\quad ({\it i\/}) &\qquad &0<N\leq {k_{\scriptstyle a}}%
\quad \mbox{and}\quad 0<N\,l\leq {k_{\scriptstyle b}}\,,  \nonumber \\
({\it ii\/}) &\qquad &0<N\leq {k_{\scriptstyle a}}\quad \mbox{and}\quad N\,l>%
{k_{\scriptstyle b}}\,,  \nonumber \\[2ex]
\lbrack {\it 2b\/}]\quad ({\it i\/}) &\qquad &N>{k_{\scriptstyle a}}\quad %
\mbox{and}\quad 0<N\,l\leq {k_{\scriptstyle b}}\,,  \nonumber \\
({\it ii\/}) &\qquad &N>{k_{\scriptstyle a}}\quad \mbox{and}\quad N\,l>{k_{%
\scriptstyle b}}\,.  \nonumber
\end{eqnarray}
Below we will examine the cases [{\it 1\/}] and [{\it 2a\/}]({\it ii\/}) in
detail as the later case contains resonances with the largest contribution
besides the required resonance at $N=0$. We have examined the other cases
and will not repeat their analysis except to note that they all give rise to
leading order corrections of order higher than those found in case [{\it 2a\/%
}]({\it ii\/}) in $\eta $. Thus, to obtain the desired Hamiltonian (\ref{140}%
), the resonances in case [{\it 2a\/}]({\it ii\/}) will impose the most
stringent condition on the size of the trap ratio $l=\nu _{\scriptstyle %
a}/\nu _{\scriptstyle b}.$ Throughout this discussion we will consider only
the lowest order terms in the Lamb-Dicke parameters, since we have already
addressed the size of the corrections to the Lamb-Dicke approximation in the
above.

In case [{\it 1\/}], we have $N=0,$ so that from the resonance condition in (%
{\ref{4020}}) we obtain $\mu-m = -{k_{\scriptstyle a}},$ and $\nu -n = - {k_{%
\scriptstyle b}}.$ This is the case which leads us to the desired
interaction ({\ref{140}}), and which we have discussed in the previous
section.

We now consider case [{\it 2a\/}]({\it ii\/}). Here, we have $0 < N \leq {k_{%
\scriptstyle a}},$ and $N\,l > {k_{\scriptstyle b}},$ so that from ({\ref
{4020}}) we obtain 
\begin{eqnarray}
m &=& \mu + {k_{\scriptstyle a}} - N \geq \mu \,,  \nonumber \\
\nu &=& n + l\,N - {k_{\scriptstyle b}} > n\,.  \label{4150}
\end{eqnarray}
Inserting these identities into ({\ref{60}}) and keeping only the lowest
order terms in the Lamb-Dicke approximation, i.e.\ $\mu = n = 0,$ we obtain 
\begin{equation}
\hat{H}_{{\rm I}} = \hbar \widetilde{g} \ \left.\!\hat{a}^\dagger\!\right.^{{%
k_{\scriptstyle a}} - N}\, \left.\!\hat{b}^\dagger\!\right.^{l\,N - {k_{%
\scriptstyle b}}} \ |1\rangle \langle3| \ + \ {\rm h.c.}\,,  \label{4160}
\end{equation}
where we have defined the coupling constant 
\begin{equation}
\widetilde{g} = -g_{13}\,\exp\left[- \frac{1}{2} ( \eta_{12}^2 + \eta_{23}^2
) \right] \frac{ (-i \eta_{12})^{{k_{\scriptstyle a}}-N} } {({k_{%
\scriptstyle a}} - N)!} \frac{(i\eta_{23})^{l\,N-{k_{\scriptstyle b}}} } {%
(l\,N - {k_{\scriptstyle b}})!}\,.  \label{4170}
\end{equation}
We require the coupling constants of the above resonances ({\ref{4150}}) to
be smaller than or equal to the coupling constants of the terms that we have
neglected in the Lamb-Dicke limit ({\ref{3990}}). Therefore, we have the
condition 
\begin{equation}
\frac{|\widetilde{g}|}{|g(0,0)|} = \frac{{k_{\scriptstyle a}} !}{({k_{%
\scriptstyle a}} - N)!} \frac{{k_{\scriptstyle b}} !}{({k_{\scriptstyle b}}
- l\,N)!} \frac{(\eta_{23})^{l N - 2{k_{\scriptstyle b}}}}{(\eta_{12})^N}
\leq \eta^2 \,,  \label{4180}
\end{equation}
where again, we assume both Lamb-Dicke parameters to be of the same order of
magnitude, $\eta_{12} \approx \eta_{23} \approx \eta.$ In order to derive a
limit for the trap ratio $l$ from the above expression, we consider the
factor containing the Lamb-Dicke parameters and the one containing the
factorials separately. If $\eta_{12} \approx \eta_{23} \approx \eta,$ we
have 
\begin{equation}
\frac{(\eta_{23})^{l N - 2{k_{\scriptstyle b}}}}{(\eta_{12})^N} \approx
\eta^{N(l-1) - 2{k_{\scriptstyle b}}} \leq \eta^2\,,  \label{4190}
\end{equation}
which is satisfied if, $N(l-1) - 2{k_{\scriptstyle b}} \geq 2.$ Since this
condition has to hold for all $N$ in the range $0 < N \leq {k_{\scriptstyle %
a}},$ this leads to the requirement 
\begin{equation}
l \geq 2{k_{\scriptstyle b}} +3 \,,  \label{4200}
\end{equation}
for the trap ratio $l.$ Next, we consider the term including the factorials.
We require this term to be smaller or equal to unity as under the above
condition ({\ref{4200}}) the factor containing the Lamb-Dicke parameters
already satisfies ({\ref{4190}}). We have 
\begin{equation}
\frac{{k_{\scriptstyle a}} !}{({k_{\scriptstyle a}} - N)!} \frac{{k_{%
\scriptstyle b}} !}{({k_{\scriptstyle b}} - l\,N)!} \leq \frac{{k_{%
\scriptstyle a}}!\,{k_{\scriptstyle b}}!}{(l\,N - {k_{\scriptstyle b}})!}
\leq \frac{({k_{\scriptstyle a}} + {k_{\scriptstyle b}})!}{(l\,N - {k_{%
\scriptstyle b}})!}\,,  \label{4210}
\end{equation}
where in the first inequality we made use of the fact that for the
resonances we are discussing here, $0 < N \leq {k_{\scriptstyle a}},$ and
the second inequality holds since, $({k_{\scriptstyle a}} + {k_{\scriptstyle %
b}})! \geq {k_{\scriptstyle a}} !\,{k_{\scriptstyle b}} !,$ for all positive
integers ${k_{\scriptstyle a}}$ and ${k_{\scriptstyle b}}.$ From the above
inequality ({\ref{4210}}) the factor containing the Lamb-Dicke parameters is
smaller or equal to unity if, $(l\,N - {k_{\scriptstyle b}})! \geq ({k_{%
\scriptstyle a}} + {k_{\scriptstyle b}})!,$ and since this has to be
satisfied for all $N$ in the range $0 < N \leq {k_{\scriptstyle a}},$ we
require 
\begin{equation}
l \geq 2 {k_{\scriptstyle b}} + {k_{\scriptstyle a}}\,.  \label{4220}
\end{equation}
Depending on the interaction that we want to generate, i.e.\ depending on
the number ${k_{\scriptstyle a}},$ the inequality ({\ref{4200}}) or ({\ref
{4220}}) will impose the stronger limit on the trap anisotropy. For the two
examples given in ({\ref{6}}) and ({\ref{7}}), we have ${k_{\scriptstyle a}}
= {k_{\scriptstyle b}} = 1,$ and ${k_{\scriptstyle a}} = 3,\,{k_{%
\scriptstyle b}} = 1,$ respectively. Therefore, in order to generate the
linear coupling Hamiltonian ({\ref{6}}) we require the trap ratio $l = \nu_{%
\scriptstyle a}/\nu_{\scriptstyle b} \geq 5,$ ({\ref{4200}}). For the cubic
interaction ({\ref{7}}) a trap ratio of $l = \nu_{\scriptstyle a}/\nu_{%
\scriptstyle b} \geq 5,$ is needed from ({\ref{4220}}).

For the remaining cases [{\it 2a\/}]({\it i\/}), [{\it 2b\/}]({\it i\/}), [%
{\it 2b\/}]({\it ii\/}) and [{\it 3\/}] a similar analysis shows that the
requirements ({\ref{4200}}) and ({\ref{4220}}) are sufficient to limit the
strength of these resonances to ({\ref{3990}}).

Although in the above discussion we have explicitly assumed the two trap
frequencies $\nu _{\scriptstyle a}$ and $\nu _{\scriptstyle b}$ to be
multiples of each other, the limits ({\ref{4200}}) and ({\ref{4220}}) also
hold for commensurate trap frequencies. In this case the trap ratio is a
rational number, i.e.\ $l=p/q,$ where $p$ and $q$ are positive integers.
Since in the resonance condition ({\ref{4020}}) all numbers need to be
integers, the number $N$ which categorizes the resonances can only take on
multiple values of $q,$ so that $l\,N=p\,N/q,$ is an integer. As we have
discussed all integer values of $N,$ any trap ratio $l=p/q,$ which satisfies
the inequalities ({\ref{4200}}) and ({\ref{4220}}) suffices for the unwanted
resonances to satisfy ({\ref{3990}}). Hence, for given values of ${k_{%
\scriptstyle a}}$ and ${k_{\scriptstyle b}}$, the coupling constants of all
additional resonances due to energy level degeneracies in the case of
commensurate trap frequencies are at least a factor of $\eta ^{2}$ smaller
than the coupling constant of the desired resonance ({\ref{140}}), if the
trap ratio is chosen large enough according to the limits in ({\ref{4200}})
and ({\ref{4220}}).

\subsection{Off-resonant terms}

\label{offresonant}

As pointed out by Gardiner {\it et.\ al.\/} \cite{gardiner97}, dropping all
off-resonant terms in going from ({\ref{90}}) to ({\ref{140}}) imposes a
limit on the time $T$ for which the Hamiltonian ({\ref{140}}) is a valid
approximation. This limit can be calculated in second order perturbation
theory to be $T\ V^2/\Delta \ll 1,$ where $V$ is the effective coupling to
the nearest off-resonant transition in ({\ref{90}}) and $\Delta$ is the
corresponding detuning. If $|M\rangle_{\scriptstyle a} |N\rangle_{%
\scriptstyle b}$ is a characteristic state which represents the highest
energy state that we allow to be acted upon, the transitions 
\begin{eqnarray}
|{M-k_{\scriptstyle a}+1}\rangle_{\scriptstyle a}|{N}\rangle_{\scriptstyle %
b} &\Leftrightarrow& |{M}\rangle_{\scriptstyle a}|{N-k_{\scriptstyle b}}%
\rangle_{\scriptstyle b} \,,  \nonumber \\
|{M-k_{\scriptstyle a}}\rangle_{\scriptstyle a}|{N}\rangle_{\scriptstyle b}
&\Leftrightarrow& |{M}\rangle_{\scriptstyle a}|{N-k_{\scriptstyle b}+1}%
\rangle_{\scriptstyle b}  \label{2050}
\end{eqnarray}
are the strongest coupled off-resonant terms. For these two transitions the
limit becomes 
\begin{eqnarray}
T\,|g(0,0)|^2 \frac{M!\,N!}{ (M-k_{\scriptstyle a}+1)!\, (N-k_{\scriptstyle %
b})! } \left(\frac{k_{\scriptstyle a}}{\eta_{12}}\right)^2 & \ll & \nu_{%
\scriptstyle a}\,,  \nonumber \\
T\,|g(0,0)|^2 \frac{M!\,N!}{ (M-k_{\scriptstyle a})!\, (N-k_{\scriptstyle %
b}+1)! } \left(\frac{k_{\scriptstyle b}}{\eta_{23}}\right)^2 & \ll & \nu_{%
\scriptstyle b}\,,  \label{2060}
\end{eqnarray}
where we have assumed the Lamb-Dicke limit to calculate the coupling $V$
between the states ({\ref{2050}}). We will further investigate the
significance of the limitations discussed here in the section below where we
concentrate on the linear coupling Hamiltonian $\hat{H}_{{\rm I}}^{{\rm (1)}%
},$ given in ({\ref{6}}).

\section{Engineering Rotation}

\label{rotation}

In the following we use the above formalism to target the linear coupling
Hamiltonian $\hat{H}_{{\rm I}}^{{\rm (1)}},$ given in ({\ref{6}}) and show
how this generates a rotation of the two-dimensional quantum motional state
of the ion. We then examine the validity of the approximations discussed in
the previous section through a numerical analysis of this specific example.

The linear coupling $\hat{H}_{{\rm I}}^{{\rm (1)}}$ is obtained from the
symmetrically combined two-mode Raman Hamiltonian ({\ref{90}}) through the
particular choice $\Delta_{13} = \nu_{\scriptstyle b} -\nu_{\scriptstyle a},$
for the Raman detuning and adjusting the relative phase of the lasers such
that the Raman coupling constant $g_{13} = i |g_{13}|,$ is purely imaginary.
This leads to the Hamiltonian 
\end{multicols}
\noindent
\rule{0.5\textwidth}{0.4pt}\rule{0.4pt}{\baselineskip}
\begin{eqnarray}
\hat{H}_{{\rm I}}^{{\rm tot}} &=& - i \hbar |g_{13}| \ \exp\left[- \frac{1}{2%
} (\eta_{12}^2 + \eta_{23}^2 ) \right] \quad \sum_{m,\mu,n,\nu} \frac{(-i
\eta_{12})^{m+\mu}}{m!\, \mu!} \frac{(i\eta_{23})^{n+\nu}} {n!\, \nu!}\,
\left.\!\hat{a}^\dagger\!\right.^m a^\mu\, \left.\!\hat{b}%
^\dagger\!\right.^\nu b^n  \nonumber \\[2ex]
&& \times\,\exp \Big [\,i\,( \nu_{\scriptstyle a} [m-\mu-1] + \nu_{%
\scriptstyle b} [\nu+1-n] + \Delta_{13})\,t\,\Big] \ + \ {\rm h.c.}\,,
\label{5990}
\end{eqnarray}
\hspace*{\fill}\rule[0.4pt]{0.4pt}{\baselineskip}%
\rule[\baselineskip]{0.5\textwidth}{0.4pt}
\begin{multicols}{2}
\noindent
which, in the limits discussed in the previous section, results in the
linear coupling $\hat{H}_{{\rm I}}^{{\rm (1)}}.$ The coupling constant $g$
in (\ref{6}) is then given by $g = - |g(0,0)|.$ The Hamiltonian $\hat{H}_{%
{\rm I}}^{{\rm (1)}}$ effects a rotation of the two-dimensional quantum
motional state of the ion about the centre of the trap. This can be seen by
examining the action of the Hamiltonian $\hat{H}_{{\rm I}}^{{\rm (1)}}$ on
the operators $\hat{a}$ and $\hat{b}.$ Using the Baker-Campbell-Hausdorff
theorem, we have 
\begin{eqnarray}
\hat{a}_{\theta} &=& \hat{U}^{(1)} \hat{a} \left.\!\hat{U}%
^{(1)}\!\right.^\dagger = \hat{a} \cos{\theta} - \hat{b} \sin{\theta}\,, 
\nonumber \\
\hat{b}_{\theta} &=& \hat{U}^{(1)} \hat{b} \left.\!\hat{U}%
^{(1)}\!\right.^\dagger = \hat{a} \sin{\theta} + \hat{b} \cos{\theta}\,,
\label{6000}
\end{eqnarray}
where the angle $\theta = g t,$ and $\hat{U}^{(1)}$ is the unitary
transformation generated by the Hamiltonian $\hat{H}_{{\rm I}}^{{\rm (1)}},$
i.e.\ 
\begin{equation}
\hat{U}^{(1)} = \,e^{i\,\hat{H}_{{\rm I}}^{{\rm (1)}} t\,/\hbar}\,.
\label{6010}
\end{equation}
From ({\ref{20}}) it is clear that the transformation in ({\ref{6000}})
corresponds to a rotation of the rescaled coordinate system $\widetilde{x}%
=x/\Delta x_0,$ and $\widetilde{y}=y/\Delta y_0,$ through an angle $\theta =
g t,$ so that in the rotated coordinate system we have 
\begin{eqnarray}
\widetilde{x}_\theta &=& \widetilde{x} \cos{\theta} - \widetilde{y} \sin{%
\theta}\,,  \nonumber \\
\widetilde{y}_\theta &=& \widetilde{x} \sin{\theta} + \widetilde{y} \cos{%
\theta}\,.  \label{6020}
\end{eqnarray}
Now an arbitrary pure or mixed motional state of the ion is characterized by
a density operator $\hat{\rho}$ which can be written as 
\begin{eqnarray}
\hat{\rho} &=& \sum_{m,n,\mu,\nu} \ \rho_{m,n}^{\mu,\nu} \ |{m}\rangle_{%
\scriptstyle a}|{n}\rangle_{\scriptstyle b} \langle{\mu}|_{\scriptstyle %
a}\langle{\nu}|_{\scriptstyle b}  \nonumber \\[1ex]
&=& \sum_{m,n,\mu,\nu} \ \rho_{m,n}^{\mu,\nu} \ \frac{\left.\!\hat{a}%
^\dagger\!\right.^m \left.\!\hat{b}^\dagger\!\right.^n}{\sqrt{m!\,n!} } \ |{0%
}\rangle_{\scriptstyle a}|{0}\rangle_{\scriptstyle b} \langle{0}|_{%
\scriptstyle a}\langle{0}|_{\scriptstyle b} \ \frac{\hat{a}^\mu \hat{b}^\nu}{%
\sqrt{\mu!\,\nu!}}\,.  \label{6030}
\end{eqnarray}
The time evolution of this state under the action of the Hamiltonian $\hat{H}%
_{{\rm I}}^{{\rm (1)}}$ is then given by 
\begin{eqnarray}
\hat{\rho}(t) &=& \hat{U}^{(1)} \hat{\rho} \left.\!\hat{U}%
^{(1)}\!\right.^\dagger  \nonumber \\[1ex]
&=& \sum_{m,n,\mu,\nu} \ \rho_{m,n}^{\mu,\nu} \ \frac{\left.\!\hat{a}%
^\dagger_{\theta}\!\right.^m \left.\!\hat{b}^\dagger_{\theta}\!\right.^n}{%
\sqrt{m!\,n!}} \ |{0}\rangle_{\scriptstyle a}|{0}\rangle_{\scriptstyle b}
\langle{0}|_{\scriptstyle a}\langle{0}|_{\scriptstyle b} \ \frac{\hat{a}%
_{\theta}^\mu \hat{b}_{\theta}^\nu}{\sqrt{\mu!\,\nu!}}  \nonumber \\[1ex]
&=& \sum_{m,n,\mu,\nu} \ \rho_{m,n}^{\mu,\nu} \ |{m}\rangle_{\scriptstyle %
a}^\theta|{n}\rangle_{\scriptstyle b}^\theta \langle{\mu}|_{\scriptstyle %
a}^\theta\langle{\nu}|_{\scriptstyle b}^\theta\,,  \label{6040}
\end{eqnarray}
where we have used (\ref{6000}) and $|{m}\rangle_{\scriptstyle a}^\theta|{n}%
\rangle_{\scriptstyle b}^\theta$ is the number state basis for the
two-dimensional harmonic oscillator but now in the rotated coordinates $%
\widetilde{x}_\theta$ and $\widetilde{y}_\theta$ as given in ({\ref{6020}}).
Therefore the motional state of the ion given by $\hat{\rho}(t)$ is
identical to $\hat{\rho},$ but rotated through an angle $\theta = gt.$ In
particular, this is accomplished without prior knowledge of the motional
state $\hat{\rho}.$

Having convinced ourselves that the linear coupling Hamiltonian $\hat{H}_{%
{\rm I}}^{{\rm (1)}}$ does rotate an arbitrary motional state of the ion we
now examine the validity of the approximations discussed in the previous
section for this specific example of the general coupling Hamiltonian ({\ref
{140}}). We consider a state rotation through the angle $\theta = \pi / 2,$
so that $T_{\rm rot} = \pi / 2 g$ is the required time to rotate the state. 
For this case, the limitations due to off-resonant terms 
(section \ref{offresonant}) as given in eq.({\ref{2060}}) take the form 
\begin{equation}
\frac{\pi}{2} N_{{\rm max}} \ll \frac{\nu_{\scriptstyle b}}{|g_{13}|}\,,
\label{6050}
\end{equation}
in the limit of small $\eta_{12} \approx \eta_{23} \approx \eta,$ and where
we have assumed $\nu_{\scriptstyle b}$ to be the smaller of the two trap
frequencies. Here, $N_{{\rm max}} = {\rm max}(N,M)$ in ({\ref{2060}}). From
eq.({\ref{6050}}) it is clear that the ratio of the lower trap frequency
over the Raman coupling constant, 
\begin{equation}
\gamma = \nu_{\scriptstyle b} / |g_{13}|\,,  \label{6055}
\end{equation}
determines the significance of off-resonant terms in the system dynamics.

From our discussion of the significance of additional on-resonant terms
(section \ref{anisotropy}) we require a trap ratio $l = \nu_{\scriptstyle a}
/ \nu_{\scriptstyle b} \geq 5,$ for the linear coupler where ${k_{%
\scriptstyle a}} = {k_{\scriptstyle b}} = 1,$ (\ref{4200}). The estimates
used to determine this minimal trap ratio essentially compare the coupling
strengths of the different terms appearing in the Hamiltonian (\ref{5990})
with no reference to the actual state on which it acts. Although this method
of estimation is used in the literature, it can only serve as a rough guide.
A more rigorous measure of how the unitary time evolution $\hat{U}^{(1)}_{%
{\rm tot}}$, generated by the symmetrically combined two-mode Raman
Hamiltonian (\ref{5990}), deviates from the desired unitary evolution $\hat{U%
}^{(1)}$, generated by the linear coupling Hamiltonian (\ref{6}), can be
quite complicated. A fully rigorous state independent measure of the
difference between two unitary operators can be constructed \cite{PERES_BOOK}%
, but we will not consider this here. In order to examine the validity of
the approximations discussed in the previous section we adopt the overlap 
\begin{equation}
\delta\equiv |\langle\Psi_{{\rm tot}}|\Psi\rangle| \,,  \label{deviation}
\end{equation}
as a measure of the deviation between the two unitary evolutions $\hat{U}%
^{(1)}_{{\rm tot}}$ and $\hat{U}^{(1)}$ for an initially pure quantum state $%
|\Psi_0\rangle.$ Here the state 
\begin{equation}
|\Psi_{{\rm tot}}\rangle = \hat{U}^{(1)}_{{\rm tot}}|\Psi_0\rangle\,,
\label{6056}
\end{equation}
gives the unitary evolution of the initial state $|\Psi_0\rangle$ under the
action of the symmetrically combined Raman Hamiltonian (\ref{5990}), and the
state 
\begin{equation}
|\Psi\rangle = \hat{U}^{(1)} |\Psi_0\rangle\,,  \label{6057}
\end{equation}
gives the desired evolution of the initial state under the action of the
linear coupling Hamiltonian (\ref{6}). This cannot be calculated
analytically. To go beyond the analytics we numerically compute the unitary
evolution (\ref{6056}), including the higher on-resonant and off-resonant
terms, on the initial pure state, $|\Psi_0\rangle=|\alpha\rangle_{%
\scriptstyle a} \otimes|\alpha\rangle_{\scriptstyle b}$, where $%
|\alpha\rangle_{\scriptstyle a}$ and $|\alpha\rangle_{\scriptstyle a}$ are
coherent states in the vibrational modes $a$ and $b$ respectively. In this
case the desired state (\ref{6057}), rotated through $\theta=\pi/2$, is
given by $|\Psi\rangle=|\! -\!\alpha\rangle_{\scriptstyle a} \otimes
|\alpha\rangle_{\scriptstyle b}.$ The results of our numerical analysis are
shown in Fig.\ref{fig5}. 
\begin{minipage}[t]{0.48\textwidth}
\begin{figure}[t]
   {   
    \leavevmode
    \epsfxsize=78mm
    \epsffile{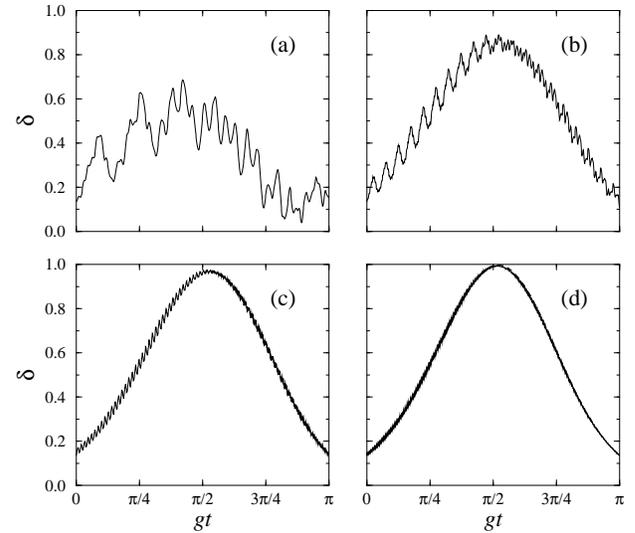}
   }   
\caption{
 Results from our numerical analysis of the deviation of the unitary
 evolution generated by the symmetrically combined Raman Hamiltonian
 tuned for rotation from the desired state rotation. We
 plot the overlap $\delta = |\langle\Psi_{\rm tot}|\Psi\rangle|$,
 between the state $|\Psi_{\rm tot}\rangle,$ resulting from the
 Raman Hamiltonian tuned for rotation and the desired state
 $|\Psi\rangle = |-\alpha\rangle_{\scriptstyle a}\otimes
 |\alpha\rangle_{\scriptstyle b},$ resulting from a rotation of the
 initial state $|\Psi_0\rangle=|\alpha\rangle_{\scriptstyle a}
 \otimes|\alpha\rangle_{\scriptstyle b},$ through the angle
 $\theta = \pi/2.$ We have chosen $\alpha = 1.$
 The graphs (a)\,--\,(d) show the dependence of
 the time evolution of $\delta$ on the parameter $\gamma$ which
 takes on the values $\gamma = 22/2^n$ where $n=3, .., 0$ in unit
 steps.
}
\label{fig5}
\end{figure}
\end{minipage}
\vspace*{\baselineskip}
 
\noindent
There we plot the overlap $\delta$ as a function of
the scaled time $gt,$ for different values of the parameter $\gamma,$ given
by (\ref{6055}) and a coherent state amplitude $\alpha=1.$ Before we discuss
our results we note the following on our choice of parameters. In Fig.\ref
{fig5}(a)\,--\,(d), $\gamma$ takes on the values $\gamma = 22/2^n,$ where $%
n=3, .., 0$ in unit steps, and the Raman coupling constant $g_{13}$ is kept
constant. For simplicity, we assume the geometry of the laser excitation to
be arranged so that the Lamb-Dicke parameters $\eta_{12} = \Delta x_0 k_{12},
$ and $\eta_{23} = \Delta y_0 k_{23},$ are equal. Here it is important to
note that the values of the Lamb-Dicke parameters $\eta_{12}$ and $\eta_{23}$
depend on the size of the trap frequencies $\nu_{\scriptstyle a}$ and $\nu_{%
\scriptstyle b}$ through $\Delta x_0 = (\hbar/2 \nu_{\scriptstyle a} m)^{1/2}
$ and $\Delta y_0 = (\hbar/2 \nu_{\scriptstyle b} m)^{1/2}.$ Therefore the
size of the Lamb-Dicke parameters depends on $\gamma$ and varies from Fig.%
\ref{fig5}(a) to (d). To incorporate this dependency in our numerical
analysis we put $\gamma \eta_{12}^2 = \gamma \eta_{23}^2 = 0.88,$ which
gives $\eta_{12} = \eta_{23} = 0.2,$ when $\gamma = 22.$ These are values
for the Lamb-Dicke parameters and the ratio $\gamma$ which have been
demonstrated in cold ion experiments \cite
{meekhof96,leibfried96,monroe96,itano97}. Following our discussion of the
trap anisotropy we choose the trap ratio $l = \nu_{\scriptstyle a}/\nu_{%
\scriptstyle b} = 5.$ Our numerical analysis was performed in a finite
(truncated) number state basis ($|{0}\rangle_{\scriptstyle a}|{0}\rangle_{%
\scriptstyle b} .. |{8}\rangle_{\scriptstyle a}|{8}\rangle_{\scriptstyle b}$%
) with a cutoff chosen such that an increase of this cutoff does not
significantly alter the result of our integration. Fig.\ref{fig5}(a) shows
the time evolution of the overlap $\delta$ for the lowest value of $\gamma.$
Here the off-resonant terms in eq.(\ref{5990}) cause strong modulations in $%
\delta.$ For higher values of $\gamma$ these modulations become much less
pronounced as the off-resonant terms contribute less on these timescales
(Fig.\ref{fig5}(b)\,--\,(d)). The time evolution of $\delta$ for the highest
value of $\gamma$ is shown with the solid line in Fig.\ref{fig5}(d). The
plot reaches a maximum of $\delta \approx 0.99$ at $g t \approx 1.02 \times
\pi/2.$ It shows almost no deviation from the dashed line in Fig.\ref{fig5}
(d) which is a numerical integration of the system dynamics where we only
include the desired resonances in the Hamiltonian as in (\ref{130}). The
numerical analysis shows that for low system excitation the Hamiltonian $%
\hat{H}_{{\rm I}}^{{\rm (1)}}$ can be engineered with high accuracy within
present ion traps. To investigate the system dynamics for higher energies
becomes computationally very expensive. To achieve the same accuracy as
obtained for $\alpha = 1$ for higher values of $\alpha,$ the number state
basis must be greatly enlarged.

In the above, we neglected decoherence. From our final comments in
section \ref{specificcoupling}, using the experimental parameters 
\cite{meekhof96} for ${}^9{\rm Be}^+$ with 
$\gamma / 2\pi = 19$MHz, $g_{13} / 2\pi = 500$kHz,
$\Delta_{12} / 2\pi = \Delta_{23} / 2\pi = 12$GHz and
$\eta_{12} = \eta_{23} = 0.2,$ we find $T_{\rm spont} \approx 200\mu s.$ 
This is to be compared with the time to rotate the motional state through
the angle $\theta = \pi/2,$ $T_{\rm rot} \approx 12\mu s.$ This confirms 
our initial assumption that decoherence through spontaneous emission
can be neglected for this process.
However, this may not be the case when engineering higher order 
interactions. One can shorten the interaction time by increasing the 
laser power while maintaining inequality (\ref{30}) by increasing the
detunings $\Delta_{12}$ and $\Delta_{23}.$ The fundamental limit is 
then given by the detunings that one can realize and the accessible
laser power.

\section{Conclusion}

In this work we showed how one can engineer a class of Hamiltonians for the
motional dynamics of an ultra-cold ion in a harmonic trap. The process uses
a stimulated Raman transition in a $\Lambda $-configuration with the two
lasers propagating along the $x$- and $y$-directions. To decouple the
internal electronic dynamics from the external motional dynamics we
constructed a Hamiltonian in which these evolutions factored. This was done
through the addition of a second pair of lasers which generated the
symmetric counter-part to the Hamiltonian generated by the first pair of
lasers. By preparing the electronic states in a particular superposition,
the internal and external dynamics completely separated and we could treat
the motional dynamics alone. In the Lamb-Dicke limit and with suitable
sideband detunings, we could ``target'' a particular term to be of leading
order in the Hamiltonian. However, we found that besides the term we wanted
to dominate, other, higher on-resonant terms appeared. We could manipulate
the strengths of the couplings to these unwanted terms by altering the trap
frequency ratio and found that we could neglect these unwanted terms in the
Lamb-Dicke approximation for large enough trap anisotropies. Finally, we did
a numerical evaluation of the full Hamiltonian as a check on the analytical
estimates. Although we have primarily concentrated on the linear rotation
Hamiltonian (\ref{6}), higher order dynamics can be generated i.e.\ $\hat{H}%
_{{\rm I}}^{{\rm (3)}},$ given by eq.(\ref{7}). The nonlinear Hamiltonian $%
\hat{H}_{{\rm I}}^{{\rm (3)}},$ has been much studied in the quantum optical
literature as a model of nonlinearly-coupled field modes \cite{drobny92}. We
know from this work that such Hamiltonians generate a rich nonlinear
dynamical structure reflecting the strong mode entanglement characteristic
of those couplings. Their optical realization is difficult, but may well be
more straightforward in trapped ion dynamics, as resonances can then be used
to isolate chosen nonlinearities. 

Finally, we note the recent publication of two papers \cite{others}
which examine types of non-linear interaction Hamiltonians in the motion
of trapped ions which are closely related to the work presented here.
\end{multicols}

\section*{Acknowledgements}

This work was supported in part by the UK Engineering and Physical Sciences
Research Council and the European Community. J.~Steinbach is supported by
the German Academic Exchange Service (DAAD-Doktorandenstipendium aus Mitteln
des zweiten Hochschulsonderprogramms). We thank Dr.~S.-C.~Gou for helpful
discussions and S.~Schneider for her helpful comments on the manuscript.

\section*{Appendix A}

\label{appenda}

In this appendix we derive the effective Hamiltonian given in ({\ref{40}})
which follows from the adiabatic elimination of the excited level $|2\rangle$
and describes the Raman coupling between the two ground state levels $%
|1\rangle$ and $|3\rangle.$ After performing the rotating wave approximation
the Hamiltonian in ({\ref{10}}) becomes 
\begin{eqnarray}
\hat{H} &=& \ \ \hbar \omega_1 |1\rangle \langle1| + \hbar \omega_2
|2\rangle \langle2| + \hbar \omega_3 |3\rangle \langle3| + \ \hbar \nu_{%
\scriptstyle a} (\left.\!\hat{a}^\dagger\!\right. \hat{a}) + \hbar \nu_{%
\scriptstyle b} (\left.\!\hat{b}^\dagger\!\right. \hat{b})  \nonumber \\[2ex]
&& - \ |1\rangle \langle2| \otimes \hbar g_{12} \,\,e^{-i\, (k_{12}\hat{x} -
\omega_{12} t)} - |2\rangle \langle1|\otimes\hbar g_{12}^* \, \,e^{i\,
(k_{12}\hat{x} - \omega_{12} t)}  \nonumber \\[2ex]
&& - \ |3\rangle \langle2|\otimes\hbar g_{23} \, \,e^{-i\, (k_{23}\hat{y} -
\omega_{23} t)} - |2\rangle \langle3|\otimes \hbar g_{23}^* \, \,e^{i\,
(k_{23}\hat{y} - \omega_{23} t)} \,,  \label{1000}
\end{eqnarray}
where we have defined the dipole coupling constants $g_{12} = \langle1| 
\underline{D}_{12}.\underline{e}_{12} |2\rangle\,E_{12} /\hbar$ and $g_{23}
= \langle3| \underline{D}_{23}.\underline{e}_{23} |2\rangle\,E_{23} /\hbar.$
In order to compare the timescales of the transitions induced by the two
laser beams we consider the Heisenberg equations of motion for the
transition operators $\hat{\sigma}_{12} \equiv |1\rangle \langle2|$ and $%
\hat{\sigma}_{13} \equiv |1\rangle \langle3|.$ 
\begin{eqnarray}
i \frac{d}{dt} \, \hat{\overline{\sigma}}_{12} &=& (\omega_2 - \omega_1) \, 
\hat{\overline{\sigma}}_{12} - g_{12}^* \,e^{i\, (k_{12}\hat{\overline{x}} -
\omega_{12} t)}\,(\hat{\overline{\sigma}}_{11} - \hat{\overline{\sigma}}%
_{22}) - g_{23}^*\,e^{i\, (k_{23}\hat{\overline{y}} - \omega_{23} t)}\,\hat{%
\overline{\sigma}}_{13}\,,  \nonumber \\[2ex]
i \frac{d}{dt} \, \hat{\overline{\sigma}}_{13} &=& (\omega_3 - \omega_1) \, 
\hat{\overline{\sigma}}_{13} - g_{23} \,e^{-i\, (k_{23}\hat{\overline{y}} -
\omega_{23} t)}\,\hat{\overline{\sigma}}_{12} + g_{12}^*\,e^{i\, (k_{12}\hat{%
\overline{x}} - \omega_{12} t)}\,\hat{\overline{\sigma}}_{23}\,.
\label{1010}
\end{eqnarray}
Here, all operators (denoted by overbars) are taken in the Heisenberg
picture, i.e.\ $\hat{\overline{\sigma}}_{12} = \hat{U} (t) \hat{\sigma}_{12} 
\hat{U}^\dagger (t),$ where $\hat{U} (t) = \hat{{\rm T}} \exp{[\,i\,(\int^t 
\hat{H} (t^\prime) dt^\prime / \hbar)\,]}$ is the time ordered evolution
operator. Using the transformation 
\begin{eqnarray}
\hat{\overline{\sigma}}_{12} &=& \,e^{-i\,\omega_{12} t}\, \hat{\widetilde{%
\sigma}}_{12}\,,  \nonumber \\
\hat{\overline{\sigma}}_{23} &=& \,e^{i\,\omega_{23} t}\, \hat{\widetilde{%
\sigma}}_{23}\,,  \nonumber \\
\hat{\overline{\sigma}}_{13} &=& \hat{\overline{\sigma}}_{12} \hat{\overline{%
\sigma}}_{23} = \,e^{-i\,(\omega_{12} - \omega_{23})t}\, \hat{\widetilde{%
\sigma}}_{13}\,,  \label{1020}
\end{eqnarray}
to remove the explicit time dependencies from ({\ref{1010}}) we have 
\begin{eqnarray}
i \frac{d}{dt} \, \hat{\widetilde{\sigma}}_{12} &=& \Delta_{12} \, \hat{%
\widetilde{\sigma}}_{12} - g_{12}^* \,e^{i\, k_{12}\hat{\overline{x}}}\,(%
\hat{\overline{\sigma}}_{11} - \hat{\overline{\sigma}}_{22}) -
g_{23}^*\,e^{i\, k_{23}\hat{\overline{y}}}\,\hat{\widetilde{\sigma}}_{13}\,,
\nonumber \\[2ex]
i \frac{d}{dt} \, \hat{\widetilde{\sigma}}_{13} &=& (\Delta_{12} -
\Delta_{23})\,\hat{\widetilde{\sigma}}_{13} - g_{23} \,e^{-i\,k_{23}\hat{%
\overline{y}}}\,\hat{\widetilde{\sigma}}_{12} + g_{12}^*\,e^{i\,k_{12}\hat{%
\overline{x}}}\,\hat{\widetilde{\sigma}}_{23}\,.  \label{1030}
\end{eqnarray}
Under the assumption of large detunings, as given in ({\ref{30}}), we obtain
the adiabatic solution for $\hat{\widetilde{\sigma}}_{12}$ by setting $d\,%
\hat{\widetilde{\sigma}}_{12}/dt \equiv 0,$ \cite{allen82} so that after
restoring the rapidly oscillating time dependence, we obtain 
\begin{equation}
\hat{\overline{\sigma}}_{12} = \frac{1}{\Delta_{12}} \left\{ g_{12}^*
\,e^{i\,(k_{12}\hat{\overline{x}} - \omega_{12} t) }\,(\hat{\overline{\sigma}%
}_{11} - \hat{\overline{\sigma}}_{22}) + g_{23}^*\,e^{i\,(k_{23}\hat{%
\overline{y}} - \omega_{23} t) }\,\hat{\overline{\sigma}}_{13} \right\}\,.
\label{1035}
\end{equation}
For the $|2\rangle \Leftrightarrow |3\rangle$ transition we find in an
analogous manner 
\begin{equation}
\hat{\overline{\sigma}}_{32} = \frac{1}{\Delta_{23}} \left\{ g_{23}^*
\,e^{i\,(k_{23}\hat{\overline{y}} - \omega_{23} t) }\,(\hat{\overline{\sigma}%
}_{33} - \hat{\overline{\sigma}}_{22}) + g_{12}^* \,e^{i\,(k_{12}\hat{%
\overline{x}} - \omega_{12} t) }\,\hat{\overline{\sigma}}_{31}\right\}\,.
\label{1040}
\end{equation}
Upon inserting these adiabatic solutions for $\hat{\overline{\sigma}}_{12}$
and $\hat{\overline{\sigma}}_{32}$ into ({\ref{1000}}), we have 
\begin{eqnarray}
\hat{H} &=& \ \ \hbar \widetilde{\omega}_1 \, |1\rangle \langle1| + \hbar 
\widetilde{\omega}_3 \, |3\rangle \langle3| + \hbar \nu_{\scriptstyle a} \,
(\left.\!\hat{a}^\dagger\!\right. \hat{a}) + \hbar \nu_{\scriptstyle b} \,
(\left.\!\hat{b}^\dagger\!\right. \hat{b})  \nonumber \\[1ex]
&& - \hbar g_{13} \, e^{-i(k_{12}\hat{x} - k_{23}\hat{y} - (\omega_{12} -
\omega_{23})t )} \,\otimes |1\rangle \langle3| - \hbar g_{13}^* \,
e^{i(k_{12}\hat{x} - k_{23}\hat{y} - (\omega_{12} - \omega_{23})t )} \,
\otimes|3\rangle \langle1| \,,  \label{1050}
\end{eqnarray}
where we have dropped the term describing the free energy of the excited
state $|2\rangle$ since in this adiabatic approximation it is no longer
connected to the two ground states. Furthermore we have defined the Raman
coupling constant as given in ({\ref{50}}), and the energies $\hbar 
\widetilde{\omega}_1$ and $\hbar \widetilde{\omega}_3$ ({\ref{55}}) of the
ground state levels $|1\rangle$ and $|3\rangle,$ which are Stark shifted as a
result of the adiabatic elimination of the excited state.

\end{document}